\documentclass[sigconf]{acmart} 

\setcopyright{acmlicensed}
\copyrightyear{2024}
\acmYear{2024}

\renewcommand\footnotetextcopyrightpermission[1]{}

\newcommand{\bench}{\textsc{RustEvo$^{\mathsf{2}}$}\xspace} 
\newcommand{\framework}{\textsc{EvoEval}\xspace}
\usepackage{float}
\usepackage{array}
\usepackage{rotating}  
\usepackage{pdflscape}
\usepackage{tabularx, multirow, makecell}

\usepackage{adjustbox}
\usepackage{amsmath,amssymb,amsfonts}
\usepackage{algorithmic}
\usepackage[linesnumbered,ruled,vlined]{algorithm2e}
\usepackage{setspace}
\usepackage{graphicx}
\usepackage{textcomp}
\usepackage{xcolor}
\usepackage{multirow}
\usepackage{mdframed}
\usepackage{balance}
\usepackage{booktabs}
\usepackage{tcolorbox}
\usepackage{threeparttable}
\usepackage{colortbl}
\usepackage{subfigure}
\usepackage{hhline}
\usepackage{pifont}
\usepackage{listings}
\usepackage{enumitem}
\usepackage{geometry}
\usepackage{verbatim}
\usepackage{xspace}
\usepackage{listings}
\usepackage{pgf-pie} 
\usepackage{pgfplots}
\usepackage{hyperref}
\usepackage{microtype}
\usepackage{subcaption}
\UseMicrotypeSet[protrusion]{basicmath}
\pgfplotsset{compat=1.16}
\newcolumntype{C}{>{\centering\arraybackslash}X}
\usepackage{pgfplots}
\pgfplotsset{compat=1.16}
\usetikzlibrary{pgfplots.statistics}

\definecolor{ggray}{HTML}{eff0f0}
\definecolor{gggray}{HTML}{E8E8E8}
\definecolor{ggggray}{HTML}{BEBEBE}

\newcommand{\et}{\textit{et al.}\xspace}

\newcommand{\claude}{Claude-3.5-Sonnet\xspace}

\definecolor{myyellow}{HTML}{FFF2CC}

\newcounter{finding}

\newcommand{\boxmargin}{1mm}
\tcbuselibrary{most, skins, breakable, theorems}
\newtcolorbox{myboxa}[2][]{
    colback=gray!10!white,
    colframe=black, enhanced,
    attach boxed title to top left={yshift=-2mm,xshift=5mm},
    title=#2,#1
}
\newtcolorbox{myboxb}[2][]{
    boxsep=3pt,
    left = \boxmargin, right = \boxmargin, top = \boxmargin, bottom = \boxmargin,
    title={#2},#1
}
\newtcolorbox{myboxc}{
    colback=gray!15!white,
    arc = 0pt, outer arc = 0pt,
    boxsep=0pt, left = 3pt, right = 0pt, top = 0pt, bottom = 0pt, 
    leftrule=3pt, bottomrule=0pt,toprule=0pt, rightrule=0pt,
    left = \boxmargin, right = \boxmargin, top = \boxmargin, bottom = \boxmargin
}
\newtcolorbox{myboxd}{
    colback=gray!10,
    colframe=black,
    width=\columnwidth,
    arc=1mm, auto outer arc,
    boxrule=0.5pt,
}

\usepackage{tikz}

\begin{document}

\title{\bench: An Evolving Benchmark for API Evolution in LLM-based Rust Code Generation}


\author{Linxi Liang\footnotemark[1]}
\email{lianglx26@mail2.sysu.edu.cn}
\affiliation{\institution{Sun Yat-sen University} \country{China}}

\author{Jing Gong\footnotemark[1] }
\email{gongj39@mail2.sysu.edu.cn}
\affiliation{\institution{Sun Yat-sen University} \country{China}}

\author{Mingwei Liu\footnotemark[2] }
\email{liumw26@mail.sysu.edu.cn}
\affiliation{\institution{Sun Yat-sen University} \country{China}}

\author{Chong Wang}
\email{chong.wang@ntu.edu.sg}
\affiliation{\institution{Nanyang Technological University} \country{Singapore}}

\author{Guangsheng Ou}
\email{ougsh3@mail2.sysu.edu.cn}
\affiliation{\institution{Sun Yat-sen University} \country{China}}

\author{Yanlin Wang}
\email{yanlin-wang@outlook.com}
\affiliation{\institution{Sun Yat-sen University} \country{China}}

\author{Xin Peng}
\email{pengxin@fudan.edu.cn}
\affiliation{\institution{Fudan University} \country{China}}

\author{Zibin Zheng}
\email{zhzibin@mail.sysu.edu.cn}
\affiliation{\institution{Sun Yat-sen University} \country{China}}

\begin{abstract}

Large Language Models (LLMs) have become pivotal tools for automating code generation in software development. However, these models face significant challenges in producing \textit{version-aware} code for rapidly evolving languages like Rust, where frequent Application Programming Interfaces (API) changes across versions lead to compatibility issues and correctness errors. Existing benchmarks lack systematic evaluation of how models navigate API transitions, relying on labor-intensive manual curation and offering limited version-specific insights. To address this gap, we present \framework, a novel framework for constructing dynamic benchmarks that evaluate LLMs' ability to adapt to evolving Rust APIs. \framework automates dataset creation by synthesizing 588 API changes (380 from Rust standard libraries, 208 from 15 third-party crates) into programming tasks mirroring real-world challenges. These tasks cover four API evolution categories — \textit{Stabilizations}, \textit{Signature Changes}, \textit{Behavioral Changes}, and \textit{Deprecations} — reflecting their actual distribution in the Rust ecosystem.

Experiments on state-of-the-art (SOTA) LLMs reveal significant performance variations: models achieve a 65.8\% average success rate on stabilized APIs but only 38.0\% on behavioral changes, highlighting difficulties in detecting semantic shifts without signature alterations. Knowledge cutoff dates strongly influence performance, with models scoring 56.1\% on before-cutoff APIs versus 32.5\% on after-cutoff tasks. Retrieval-Augmented Generation (RAG) mitigates this gap, improving success rates by 13.5\% on average for APIs released after model training. Our findings underscore the necessity of our evolution-aware benchmarks \bench to advance LLMs’ adaptability in fast-paced software ecosystems. The framework and dataset are publicly released at~\cite{RustAPIEvo}.

\end{abstract}


\maketitle

\footnotetext[1]{Equal contribution.}
\footnotetext[2]{Corresponding author.}  



\section{Introduction}


LLMs for code generation have become powerful tools for enhancing developer productivity, assisting in writing code across diverse programming languages\cite{cao2025buildbenchmarkrevisiting274, zheng2024understandinglargelanguagemodels, 10006873, zhuo2024bigcodebenchbenchmarkingcodegeneration, tian2023chatgptultimateprogrammingassistant, jin2024llmsllmbasedagentssoftware}. However, these models face significant challenges in maintaining knowledge timeliness, particularly when generating code for languages with evolving  APIs~\cite{wu2024versicode, chen2021evaluating, 8530111, 10.1145/3338906.3338971, mahmud2024automatedupdateandroiddeprecated, sun2022miningandroidapiusage}. Programming languages and their ecosystems continuously evolve to introduce new features, improve performance, and enhance security. Rust, known for its strong emphasis on memory safety and performance, has undergone particularly rapid evolution, with 85 release versions since its 1.0 launch in 2015. Different types of API changes, such as newly stabilized features and behavioral adjustments, are introduced with each Rust release. Despite advances in LLM-based code generation, there remains a critical gap in systematically understanding how effectively these models navigate such rapid API evolution, especially within Rust's unique versioning model and evolution patterns~\cite{mcnamara2021rust, klabnik2019rust}.



Recently, several studies have investigated LLMs' ability to handle API evolution, particularly in established languages like Python. Wang \et~\cite{wang2025llmsmeetlibraryevolution} conduct an empirical study showing that LLMs frequently use deprecated Python APIs in code completion tasks, highlighting their difficulty in consistently adopting up-to-date APIs. Wu \et~\cite{wu2024versicode} evaluate LLMs' capabilities in code generation under version-specific library requirements, revealing significant performance degradation when the required API versions differ from the predominant ones in their training data. Liu \et~\cite{liu2024codeupdatearena} assess LLMs' adaptability in API updates in code generation, demonstrating that RAG with change documentation can be a promising approach to mitigating challenges posed by API evolution in LLM-based code generation.

\begin{figure}
    \centering
    \includegraphics[width=1.0\linewidth]{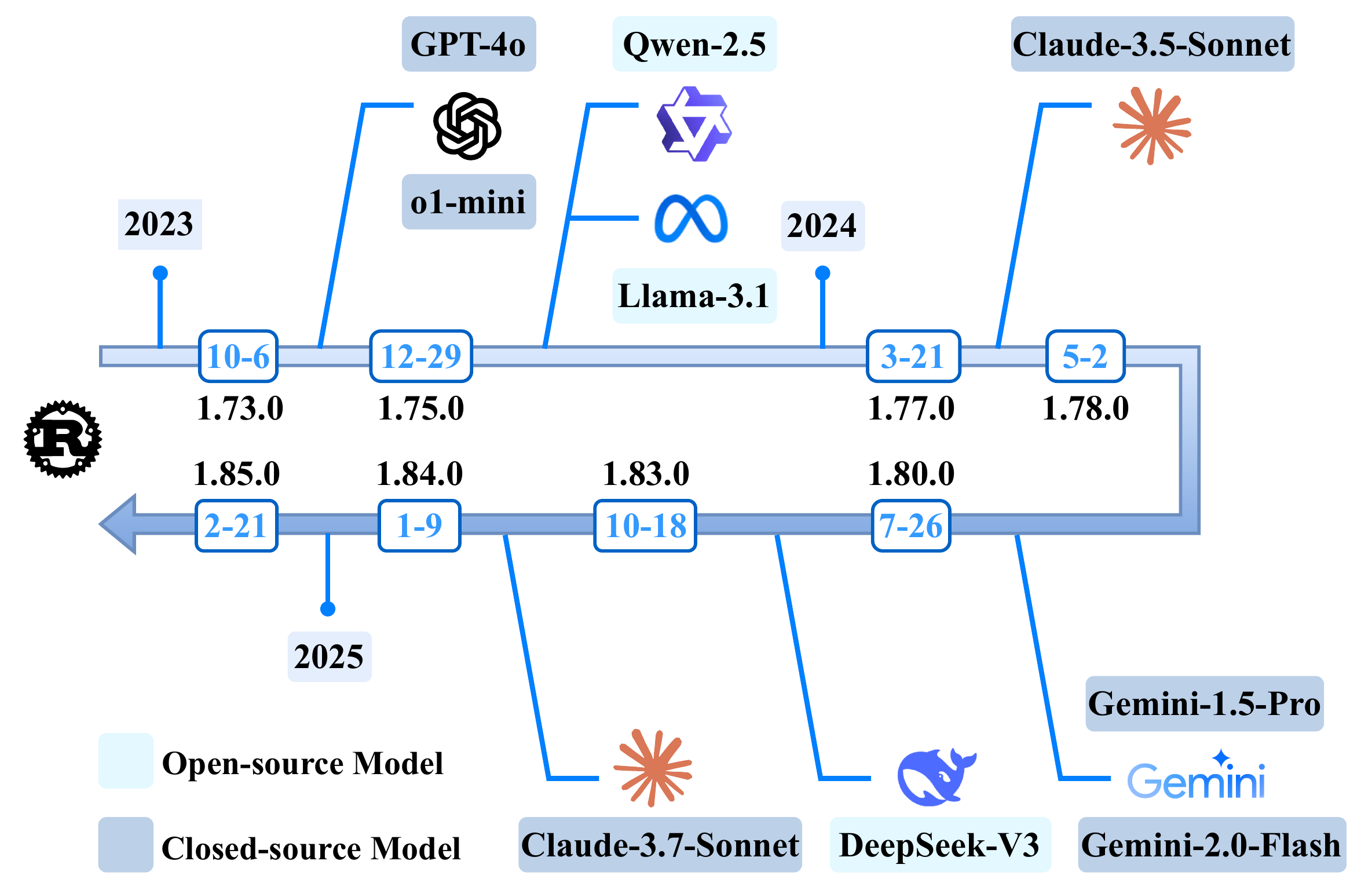}
    \caption{Timeline illustrating the release dates of Rust versions and knowledge cutoff points of LLMs (2023–2025).}
    \label{fig:timeline}
\end{figure}

Although these existing studies have provided valuable insights, their methodologies and findings face the following limitations when applied to rapidly evolving languages like Rust.
\begin{enumerate}[label=\textbf{C\arabic*}, leftmargin=15pt, labelsep=0.5em, itemsep=0.5em, topsep=0.5em]
    \item \textbf{LLMs Are More Likely to Face Difficulty with Rapidly Evolving APIs.} Due to Rust's rapid version evolution, its APIs frequently change for various reasons. For example, our preliminary statistics indicate that at least 495 API changes have occurred in Rust over 29 months, across 15 version updates, from July 2023 to February 2025. This high frequency of changes results in sparse or unbalanced occurrences of certain Rust API versions in open-source repositories. As a result, LLMs trained on such data are \textit{more likely} to struggle with accurately using the intended Rust API versions compared to other established languages like Python. Wang \et demonstrate that the rapid evolution of APIs leads to more frequent use of deprecated APIs by LLMs. This finding is supported by a comparative analysis of Python libraries with varying rates of evolution ~\cite{wang2025llmsmeetlibraryevolution}. While Python releases a new major version, such as transitioning from 3.12 to 3.13, approximately once a year~\cite{pythonversions}, Rust evolves much more rapidly, with a new major version, like 1.84 to 1.85, being released approximately every six weeks~\cite{rustchannels}. Therefore, a separate evaluation of API evolution in LLM-based Rust code generation is essential for further exploration.

    \item \textbf{Timeliness of Evaluation Benchmarks Is Limited Due to Rapid API Evolution.} LLMs typically have a knowledge cutoff date, meaning they are trained on data collected before that time. Benchmarks used to evaluate the code generation capabilities of LLMs also have their own knowledge cutoff dates. However, as Rust versions evolve and new LLMs are released, these benchmarks can become \textit{outdated} and no longer accurately reflect the current capabilities of LLMs with evolving APIs. As illustrated in Figure~\ref{fig:timeline}, Rust versions continue to evolve alongside LLM update cutoff dates. If an evaluation benchmark is statically constructed based on a specific snapshot of Rust APIs, the resulting evaluation likely become outdated and provide misleading insights into the latest Rust version.
\end{enumerate}

To address these limitations, we introduce \textbf{\framework}, a novel framework for continuously constructing and \textbf{evo}lving benchmarks to \textbf{eval}uate the ability of LLM-based code generation with Rust API evolution. Note that, although our study focuses on the rapidly evolving Rust, our framework is inherently language-agnostic.
The framework operates in two automated phases. In the first phase, it continuously gathers API changes from multiple sources alongside Rust version updates, including official changelogs, documentation differences, and repository implementation changes. These changes are classified into four categories: Stabilizations, Signature Changes, Behavioral Changes, and Deprecations.

In the second phase, the collected API evolution changes are automatically transformed into synthetic programming tasks. Specifically, for each API change, we leverage LLMs guided by carefully designed prompts tailored to different change categories to automatically generate programming tasks. Each task includes a natural language description, a reference solution, and executable test programs. These tasks implicitly require the use of specific API versions while reflecting realistic programming scenarios. 

We apply our framework to recent Rust versions (1.71.0-1.84.0) and construct a dataset containing 588 API changes, with 380 (64.6\%) from the official Rust language libraries and 208 (35.4\%) from 15 popular third-party crates. The resulting programming tasks cover all four categories of API changes, with the distribution reflecting the actual frequency of different change types in the ecosystem (31.3\% stabilizations, 31.5\% signature changes, 33.2\% behavioral changes, and 4.1\% deprecations).
We evaluate these tasks using ten SOTA language models and found significant performance variations based on model knowledge cutoff dates and API change types. Models perform better on tasks involving stabilized APIs (average 65.8\% success rate) compared to behavioral changes (38.0\%), highlighting the challenge of detecting behavioral modifications without signature changes. Additionally, our experiments with RAG demonstrate substantial improvements (13.5\% average increase) in handling API changes released after model knowledge cutoff dates. 

Our main contributions are:
\begin{itemize}[leftmargin=15pt, labelsep=0.5em, itemsep=0.5em, topsep=0.5em] 
    \item We propose a two-phase framework, \framework, for constructing Rust API evolution datasets that combine multi-source data collection with LLM-based task generation. Our framework then leverages LLMs to transform this raw evolution data into natural programming tasks that implicitly require specific API versions.
    \item We construct \bench, a dataset containing 588 API changes from recent Rust versions paired with programming tasks designed to evaluate API-aware code generation capabilities. 
    \item We conduct experiments evaluating how SOTA LLMs perform on API evolution tasks, revealing significant performance variations based on model knowledge cutoff dates and API change types.
    We provide the code and data at~\cite{RustAPIEvo}.
\end{itemize}




\section{BACKGROUND AND MOTIVATION}

\subsection{Version-Aware Code Generation}
Version-aware code generation tackles the challenge of producing code compatible with specific library or framework versions in evolving software ecosystems, where deprecated APIs or syntax changes break functionality. Traditional models trained on aggregated datasets without version metadata struggle with this temporal sensitivity ~\cite{allal2023santacoder}. Recent efforts address this by curating version-labeled datasets and incorporating version constraints. For instance, Wu \et introduce VersiCode, a dataset for evaluating version-specific code generation, and a retrieval-augmented approach that uses version-specific documentation and API references ~\cite{wu2024versicode}.


A key enhancement to version-aware code generation comes from the integration of RAG ~\cite{du2024vulragenhancingllmbasedvulnerability}, which dynamically retrieves version-specific documentation, or code examples to guide the generation process. This approach overcomes the limitations of traditional LLMs, which rely solely on parametric knowledge that may not reflect the latest library updates. The synergy between RAG and version-aware code generation is particularly valuable in continuous integration and continuous deployment (CI/CD) workflows, where maintaining compatibility across versions is essential to avoiding deployment disruptions ~\cite{thongtanunam2016revisiting}. 

Despite these advancements, version-aware code generation faces ongoing challenges. The rapid evolution of software ecosystems demands continuously updated datasets to capture the latest library versions. Additionally, efficient retrieval mechanisms are necessary to navigate the vast and ever-growing array of library versions, particularly in real-time development tools where latency and accuracy must be carefully balanced. While RAG offers significant benefits, its integration introduces computational overhead, impacting latency and resource use, which must be optimized for practical deployment. Nevertheless, the ability of version-aware code generation to produce robust, compatible codebases in dynamic software environments positions it as an indispensable tool for addressing the complexities of modern software development.

\subsection{Retrieval-Augmented Generation}


In code generation, RAG typically involves two components: (1) a Retriever that fetches relevant content using techniques like dense vector search (e.g., FAISS ~\cite{johnson2019billion}) or sparse retrieval (e.g., BM25 ~\cite{robertson2009probabilistic}); and (2) a Generator, such as GPT or CodeBERT ~\cite{feng2020codebert}, that synthesizes the final output by integrating the retrieved information with the input query ~\cite{lewis2020retrieval}. 
The effectiveness of RAG is demonstrated by frameworks such as REDCODER ~\cite{parvez2021retrieval}, which accesses code and summaries from unimodal and bimodal databases. Evaluations on datasets like CONCODE ~\cite{iyer2018mapping}, CodeSearchNet ~\cite{husain2019codesearchnet}, and CodeXGLUE ~\cite{lu2021codexglue} show that REDCODER outperforms traditional models by mimicking developers’ practice of reusing past code, thus enhancing both accuracy and relevance.

However, RAG faces a challenge. Beyond computational overhead, the quality of output depends on the relevance of retrieved data. Poorly curated sources can introduce errors, prompting ongoing research into optimizing retrieval and integration techniques ~\cite{izacard2020leveraging}. Despite these hurdles, RAG’s applications in code completion, bug fixing, and documentation generation make it a versatile tool for modern development ~\cite{parvez2021retrieval}.

\begin{figure}
    \centering
    \includegraphics[width=1.0\linewidth]{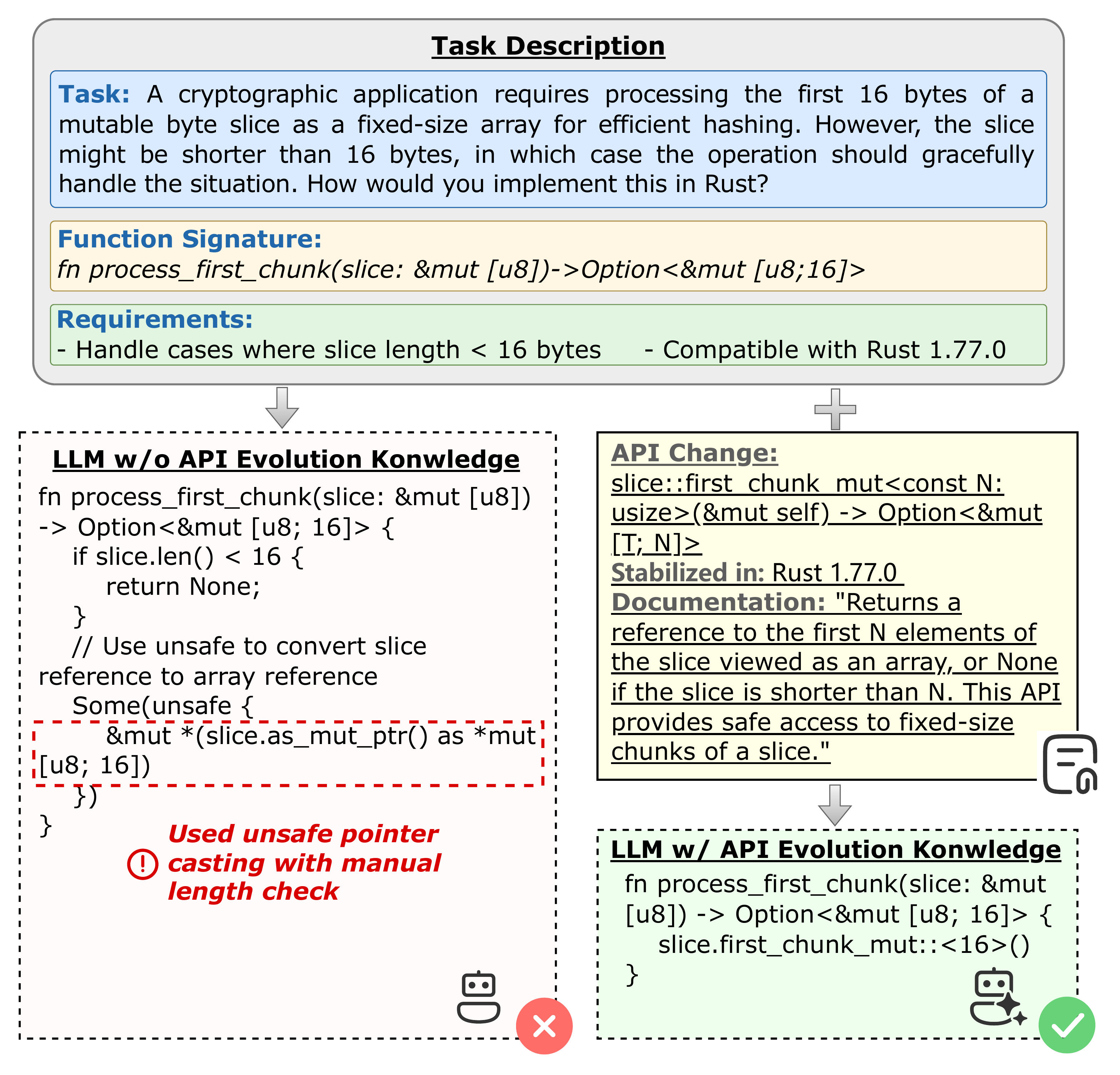}
    \caption{A Motivating Example}
    \label{fig:motivation}
    \vspace{-10pt}
\end{figure}

\subsection{A Motivating Example}
To illustrate the challenges of code generation in the context of evolving APIs, we consider a real-world case from the Rust ecosystem. In Rust 1.77.0 (released in March 2024), the \texttt{slice::first\_chunk\_mut} method\footnote{\url{https://doc.rust-lang.org/stable/std/primitive.slice.html\#method.first\_chunk\_mut}} was stabilized, providing a safer and more efficient way to obtain a reference to the first N elements of a slice as a fixed-size array.  
Figure~\ref{fig:motivation} presents a programming task that implicitly requires using an API that was stabilized well before the model’s knowledge cutoff date. However, when prompted with this task, \claude~\cite{anthropic2024claude3} — a SOTA LLM with a knowledge cutoff in April 2024~\cite{anthropicDocs} — still generates code relying on an older, less safe pattern involving unsafe pointer casting and manual length checking. While this approach remains functional, it introduces unnecessary unsafe code and forgoes the benefits of improved safety and readability. In contrast, when explicitly informed about the API evolution, the same model correctly utilizes \texttt{slice::first\_chunk\_mut}, producing more concise and robust code.  

This example highlights a fundamental challenge in API-aware code generation: \textbf{LLMs often default to historical coding patterns rather than leveraging modern APIs, even when the newer APIs were available before their knowledge cutoff.} This issue arises due to two factors: (1) Persistence of Legacy Idioms. Developers tend to prefer older, more familiar practices even when newer and better alternatives are available. This preference is driven by familiarity, comfort, and the perceived stability of legacy approaches ~\cite{talejko1967principal}. (2) Training Data Bias. Without explicit version-specific guidance, LLMs tend to favor patterns they encountered more frequently, potentially overlooking API improvements.


This challenge underscores the necessity of a dedicated benchmark to assess LLMs' ability to adapt to API evolution in real-world coding scenarios. \textbf{Existing code generation benchmarks often fail to systematically evaluate whether models can utilize the latest API features, that API evolution knowledge, instead measuring performance on static datasets that do not reflect ongoing language and ecosystem changes.} A static benchmark is insufficient for evaluating API-aware code generation, as it cannot account for the continuous evolution of programming languages, libraries, and best practices.  

To address this limitation, \textbf{our benchmark is designed to evolve alongside API changes, ensuring that evaluation remains relevant as new APIs are introduced and old patterns become obsolete.} Rather than relying on manually curated tasks, our dataset construction process is designed to be \textbf{automated}, dynamically incorporating API updates to create new evaluation instances. This automation enables continuous and scalable assessment of LLMs' ability to generate optimal, up-to-date code.  


\section{Approach}
\label{sec:app}
To bridge the gap in existing benchmarks, we propose \framework, a two-phase framework for continuously constructing \bench to track Rust API evolution. By integrating API changes, \bench enables dynamic and scalable evaluation of LLMs' ability to generate code with the latest standards.

\subsection{Overview}
\begin{figure*}
    \centering
    \includegraphics[width=1.0\linewidth]{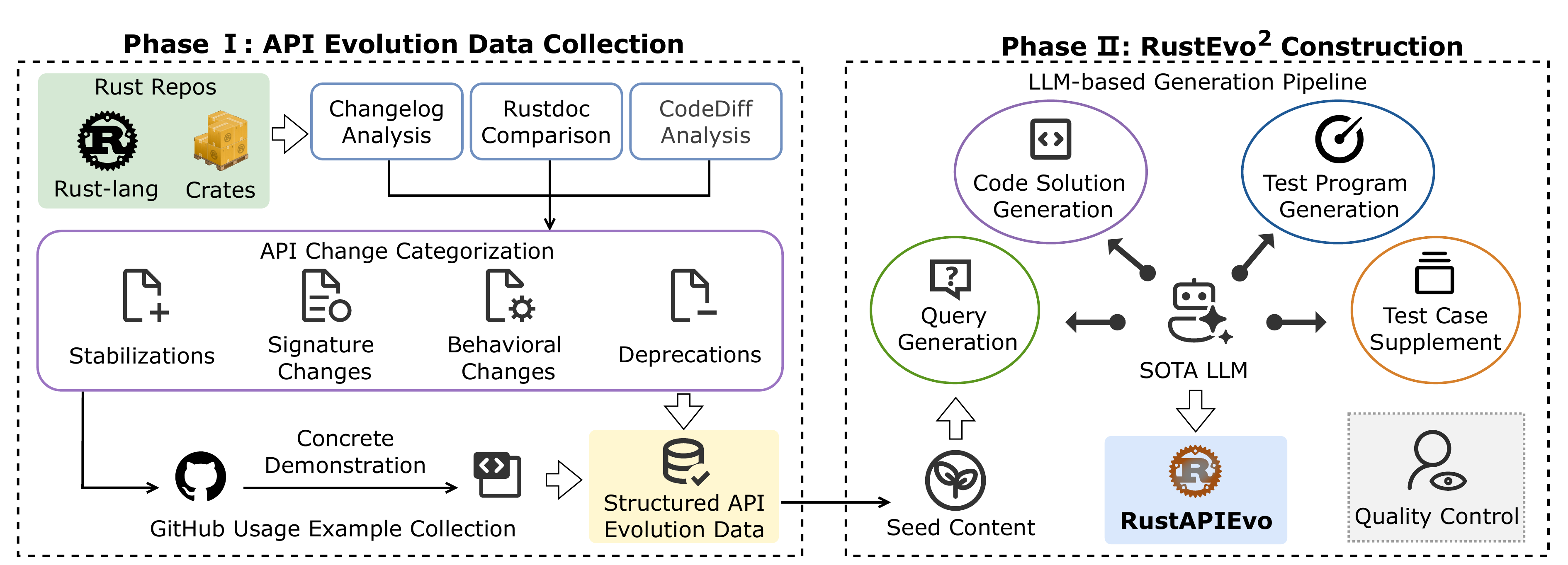}
    \caption{Two-phase \bench Construction Framework}
    \label{fig:overview}
\end{figure*}

The \bench construction framework, illustrated in Figure \ref{fig:overview}, consists of two key phases: \textbf{API Evolution Data Collection} and \textbf{\bench Construction}.

In the first phase, our primary objective is to collect detailed information on Rust API evolution from diverse sources. We analyze official Rust repositories and third-party crates to identify API changes across recent versions. The collection process involves extracting changelogs, comparing differences in API documentation, and analyzing alterations in code implementations. We then mine GitHub for real-world code examples that utilize these APIs. This phase produces a structured documentation of API evolution, categorizing changes into four types based on their evolution patterns and how they affect developers' existing code: Stabilizations, Signature changes, Behavioral Changes, and Deprecations.

In the second phase, we use the collected API evolution data to generate programming tasks for evaluation. We curate seed content from the API documentation and GitHub usage examples. Then, a LLM-based generation pipeline is employed to automatically produce natural programming queries, corresponding code solutions, and test programs. Each generated task undergoes quality control to verify its correctness and relevance. The final output is a set of programming tasks, each accompanied by test cases capable of distinguishing between legacy and updated API usage.
The two phases are designed to work sequentially, with the API evolution data from Phase I serving as input to the automatic task generation process in Phase II. This approach allows for focused data collection followed by targeted task construction that reflects real-world programming scenarios involving API evolution. Although our research primarily centers on Rust, the two-phase framework can be readily adapted to assess API evolution in other programming languages.

\subsection{Phase \uppercase\expandafter{\romannumeral 1}: API Evolution Data Collection}

In this phase, we gather API evolution information from multiple sources and process it to create a detailed record of how Rust APIs have changed across versions. Our approach focuses on recent Rust releases to capture the most relevant API changes for developers.

\subsubsection{Data Sources}
Our data collection process draws from two primary sources: the official Rust repository and selected third-party crates. For each source, we apply specialized analysis techniques to identify and document API changes.
From the official Rust repository, we extract API evolution information through three complementary methods:

(1) We perform changelog analysis to identify explicitly documented changes, focusing on API stabilizations and deprecations announced in official release notes.

(2) We conduct Rustdoc documentation~\cite{docsrs} comparisons to detect changes in signatures and documented behaviors, using Python-based \texttt{BeautifulSoup} and \texttt{difflib} libraries to parse and compare HTML documentation between successive Rust releases.

(3) We directly analyze source code from the official Rust repository to identify functional changes that may not be explicitly documented. This includes detecting Rust attributes like \texttt{\#[deprecated()]} and \texttt{\#[stable()]}, as well as comparing implementation details to find behavioral changes.


For third-party crates, we select 15 candidates based on quantitative criteria such as star count, update frequency and documentation completeness, focusing on widely used libraries with adequate documentation to track API evolution. After selection, we apply similar code implementation change tracking techniques to identify evolution patterns in these libraries. This multi-method approach captures a broader range of API changes than any single analysis technique, particularly for detecting implicit functional changes that do not alter API signatures.

\subsubsection{API Change Categorization}

Based on our analysis, we categorize API changes into four distinct types that represent different evolution patterns in the Rust ecosystem:
\begin{itemize}
    \item \textbf{Stabilizations} refer to newly introduced transitions of APIs from unstable to stable status, signifying functionality that is now officially supported for production use.
    \item \textbf{Signature Changes} modify the interface contract while preserving core functionality. These changes encompass modifications to parameters, return types, and generic type constraints (e.g., trait bounds, associated type bounds, and where clauses), which necessitate adaptations in client code. 
    \item \textbf{Behavioral Changes} affect the underlying implementation behavior without altering the API signature, and their identification relies on analyzing both code differences and corresponding documentation updates that indicate behavioral modifications.
    \item \textbf{Deprecations} denote functionality that is explicitly marked for removal or substitution, typically with migration guidance towards alternative implementations. 
\end{itemize}
This categorization provides a systematic framework for analyzing various API evolution events and assessing their potential impact on dependent code.

\subsubsection{GitHub Usage Example Collection}
To provide realistic seed content for LLM-based task generation in Phase \uppercase\expandafter{\romannumeral 2}, we systematically collect real-world code examples that utilize the collected evolving APIs. These examples serve as concrete demonstrations of API usage patterns, enabling LLMs to generate natural programming tasks based on authentic coding scenarios.

\textbf{Usage Example Extraction.}
To enhance the relevance and accuracy of the collected code examples, we follow a systematic approach: (1) \textbf{Repository Selection.} We crawl GitHub repositories that have been updated after the release of specific Rust versions. (2) \textbf{Code Segment Identification.} Within these repositories, we identify code segments that utilize the APIs documented in our evolution dataset. (3) \textbf{Temporal Filtering.} By applying temporal filtering, we verify that the collected examples are relevant to the specific API versions under study. 


\textbf{Contextual Information Extraction.} For each API usage instance, we extract not only the direct API call but also 20 lines of surrounding context, including any relevant variable declarations, function calls, or other related operations that may influence the behavior or understanding of the API usage. This contextual information is crucial as it captures the broader programming scenario in which the API is used. When provided to LLMs in Phase II, this context enables generation of more realistic and nuanced programming tasks that reflect authentic developer challenges.

\textbf{API Version Verification.} A key technical challenge in this collection process is accurately determining which version of an API a code example utilizes. To address this challenge, we implement a multi-faceted verification approach that combines static analysis using the tree-sitter library~\cite{treesitter} to parse and analyze code structures and API references. This static analysis is complemented by examining \texttt{cargo.toml} manifests to identify declared dependency versions and further validated through compilation testing with version-specific Rust toolchains. This integrated approach results in more reliable attribution of code examples to specific API versions, allowing the collected examples to accurately reflect the evolution patterns captured in our dataset.

\subsection{Phase \uppercase\expandafter{\romannumeral 2}: \bench Construction}
The second phase of our framework focuses on transforming the API evolution data collected in Phase \uppercase\expandafter{\romannumeral 1} into realistic programming tasks. This phase utilizes the collected API documentation and usage examples as seed content for an LLM-based generation pipeline.
\subsubsection{Seed Content Preparation}

We prepare seed content using two primary sources collected in Phase \uppercase\expandafter{\romannumeral 1}: API documentation and real-world usage examples. For each API in our evolution dataset, we extract signature information (including names, parameters, and return types), documentation that describes functionality and usage, usage examples sourced from both official documentation and GitHub repositories, and contextual details regarding version-specific behavior. This information is then structured into a format suitable for input to LLMs. For APIs with multiple versions, we maintain separate seed content for the pre-change and post-change states, thereby enabling the generation of tasks that assess migration capabilities across different versions.

\subsubsection{LLM-based Generation Pipeline}

\begin{figure*}
    \centering
    \includegraphics[width=1\linewidth]{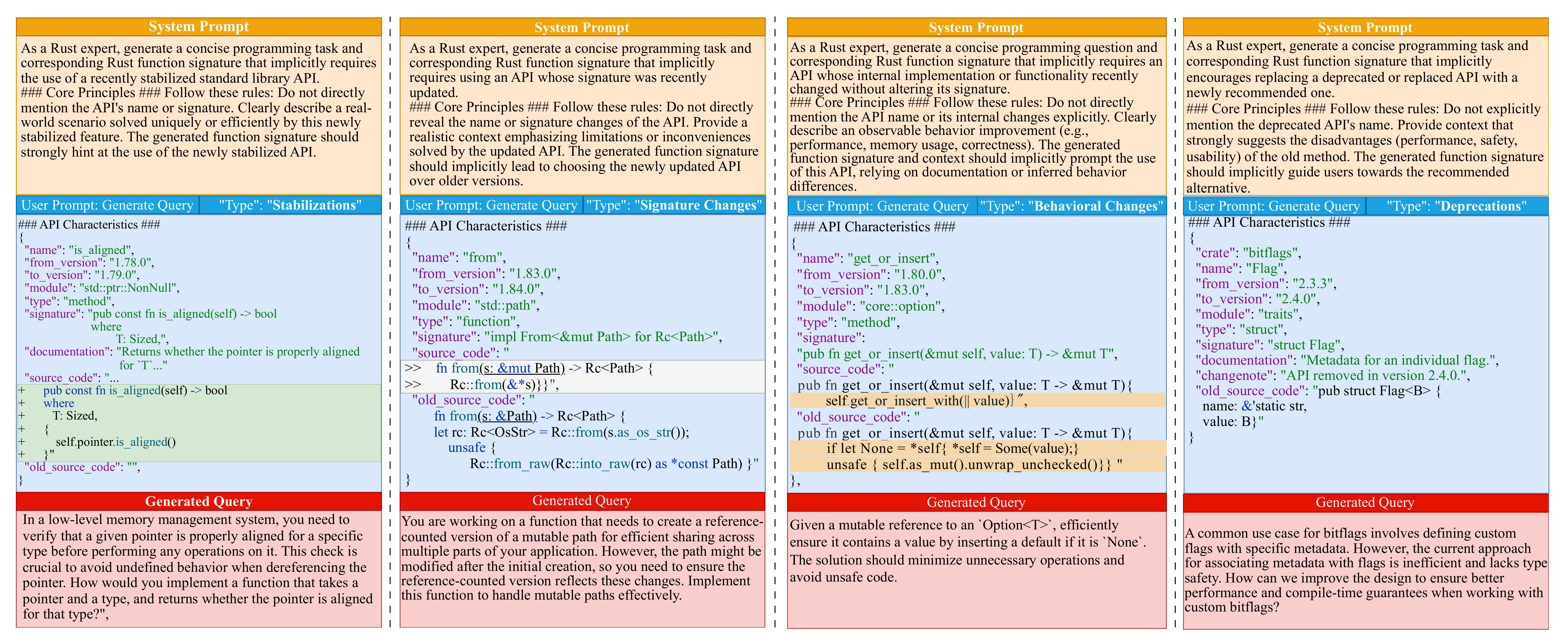}
    \caption{Query and signature generation examples corresponding to four API change categories: Stabilizations, Signature Changes, Behavioral Changes, and Deprecations.}
    \label{fig:categories_examples}
\end{figure*}

We develop a multi-stage generation pipeline to create programming tasks based on API evolution data. This pipeline consists of four sequential stages, each designed to address a specific aspect of task creation:

\textbf{Query Generation.} The first stage produces natural language programming queries that implicitly require the use of target APIs. We prompt LLMs with API documentation and contextual information, instructing it to create concise tasks that focus on one core problem without explicitly mentioning the target API. As shown in Figure~\ref{fig:categories_examples}, our prompts vary based on the evolution pattern: Stabilization prompts focus on unique solutions offered by new features; Signature Change prompts focus on the updated parameter structures or return types; Behavioral Change prompts highlight observable changes in functionality without signature alterations; and Deprecation prompts guide towards recommended alternatives without naming deprecated APIs. This approach yields tasks that test API knowledge organically rather than through direct instruction.

\textbf{Code Solution Generation.} Using the generated queries and API information as input, the second stage creates reference implementations that correctly utilize the target APIs. These solutions serve as ground truth for evaluating model-generated code in downstream assessments.

\textbf{Test Program Generation.} The third stage constructs executable test programs tailored to each task. These include two checks: static analysis checks that confirm the presence of required APIs in submitted solutions, and dynamic test cases that verify functional correctness across various inputs.

\textbf{Test Case Augmentation.} The final stage expands the initial test cases to improve coverage and enhance the test program's ability to differentiate between old and new API usage patterns. This stage focuses on creating test cases that can detect subtle behavioral differences in API implementations across versions. Our approach generates a suite of 2,574 test cases across all programming tasks, with an average of 4.38 test cases per test program, providing thorough evaluation of API usage across diverse scenarios.

We maintain atomicity in the generation process by prompting the model separately for each stage, passing only essential context between steps. This minimizes error propagation and enables targeted refinement at each stage. Our implementation uses DeepSeek-v3~\cite{liu2024deepseekv3}, known for its high performance and cost efficiency. The pipeline delivers high-quality results, with 85.7\% (588/684) of initially generated test programs successfully passing compilation and execution tests without manual intervention. We retain only these 588 validated examples in our final dataset, filtering out the 96 that failed due to compatibility issues or missing dependencies.

\begin{figure}
    \vspace{-5pt} 
    \centering
    \includegraphics[width=1.0\linewidth]{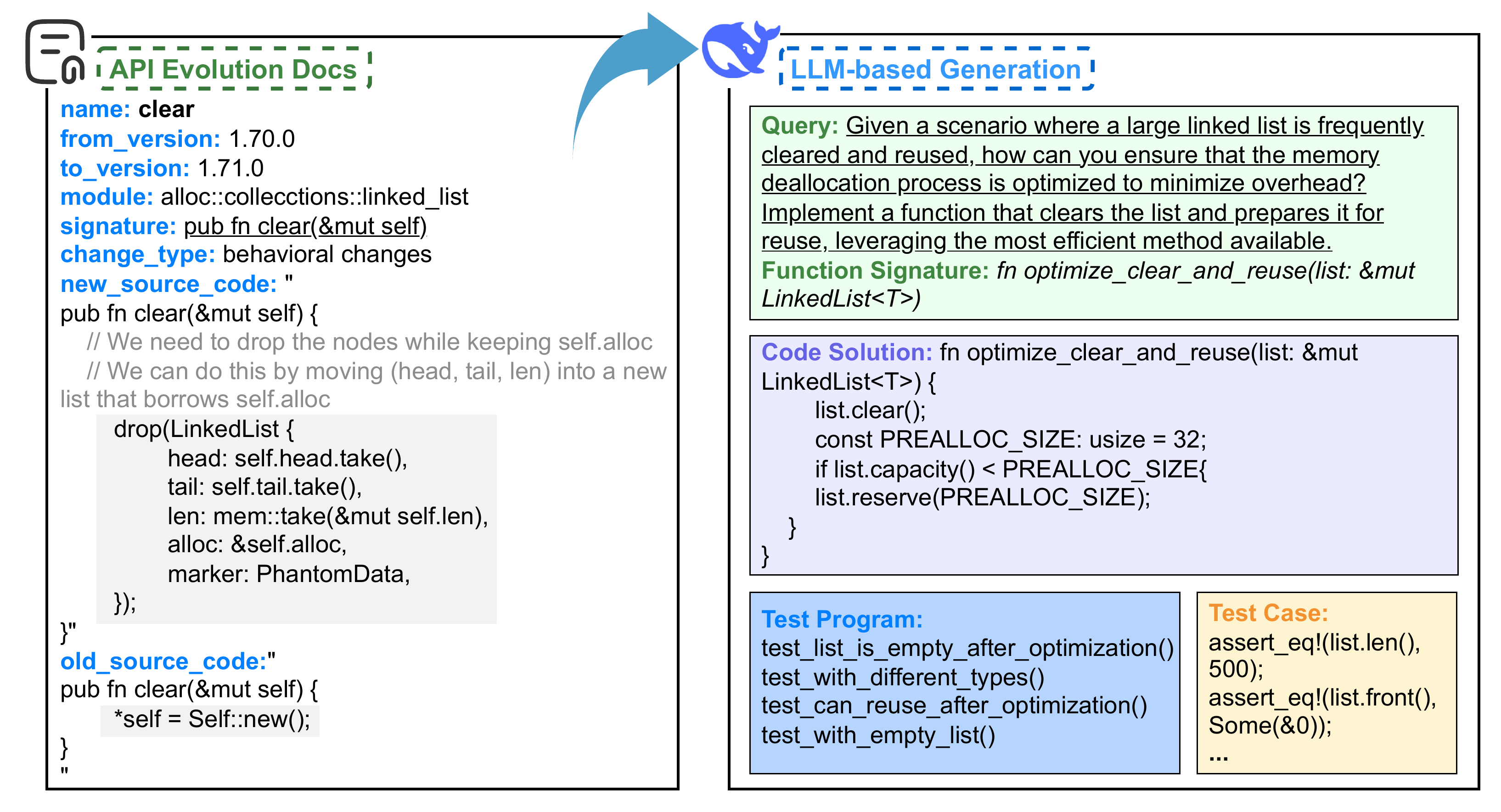}
    \caption{The \framework pipeline illustrated through an example.}
    \label{fig:example}
    \vspace{-10pt}
\end{figure}

Figure~\ref{fig:example} illustrates this pipeline with an example task involving \texttt{LinkedList} memory management optimization, showing the progression from initial API documentation to the final programming task with test cases. The staged approach allows for quality control at each step, with the option to regenerate specific components if they fail to meet quality standards.


\subsubsection{Quality Control}

To maintain the quality of the generated tasks, we implement a series of verification mechanisms. A judge model reviews each task to detect issues such as the explicit mention of target APIs in queries, misalignment between the queries and their corresponding solutions, technical errors in test programs, and insufficient test coverage. Tasks that do not meet these standards are either regenerated or refined through additional prompting. In addition, automated verification procedures assess the technical correctness of the generated code, including compilation testing with the appropriate Rust versions.

\section{Experimental Setup}
We introduce the research questions, the basic experimental setup about the datasets and models, and the evaluation metrics used during the evaluation. The research questions include:

\textbf{RQ1: How do different models perform on our \bench?} We compare the performance of various SOTA LLMs on our \bench benchmark. For each model, we provide corresponding API information and measure its ability to generate correct Rust code that properly utilizes the specified APIs. We analyze performance differences between models to identify key factors influencing API-aware code generation capability. 

\textbf{RQ2: How do models perform across different API change types?} We analyze model performance across the four categories of API changes (stabilized APIs, signature changes, implicit functional changes, and deprecated/replaced APIs). For each category, we measure success rates and identify specific patterns of errors. This analysis helps determine which types of API changes present the greatest challenges for LLMs and informs the development of targeted solutions.

\textbf{RQ3: What is the relationship between models' knowledge cutoff dates and their performance on API evolution tasks?} We investigate how a model's knowledge cutoff date affects its ability to handle API changes released at different points in time. We group tasks based on whether the API changes occurred before or after each model's knowledge cutoff date and compare performance across these temporal groups. This reveals how well models can adapt to previously unseen API changes and informs strategies for maintaining model relevancy in rapidly evolving Rust ecosystems.
    
\textbf{RQ4: How effective are RAG methods in improving performance on API evolution tasks} We evaluate three approaches for handling API evolution tasks: (1) baseline performance without additional information, (2) RAG using our API evolution documentation, and (3) direct provision of ground truth API information. We focus particularly on API changes released after model knowledge cutoff dates to determine how effectively different approaches can bridge knowledge gaps.


\textbf{RQ5: How does model performance on evolving APIs compare to performance on stable (non-evolving) APIs?} 
We create a control dataset containing 50 stable APIs from the official Rust repository that were stabilized prior to Rust 1.70.0 (January 2023) and have not undergone any breaking changes since. Using the same \framework method, we construct programming tasks for these stable APIs. We then select 50 comparable tasks from \bench (also derived from the official Rust libraries) for direct comparison. Both sets are evaluated using identical metrics (Pass@1) and under the same conditions (providing API information during testing). This controlled comparison quantifies the specific challenges introduced by API evolution while maintaining consistency in API source, task construction methodology and evaluation approach.

\subsection{Dataset Statistics}

\begin{figure}
    \centering
    \includegraphics[width=1.0\linewidth]{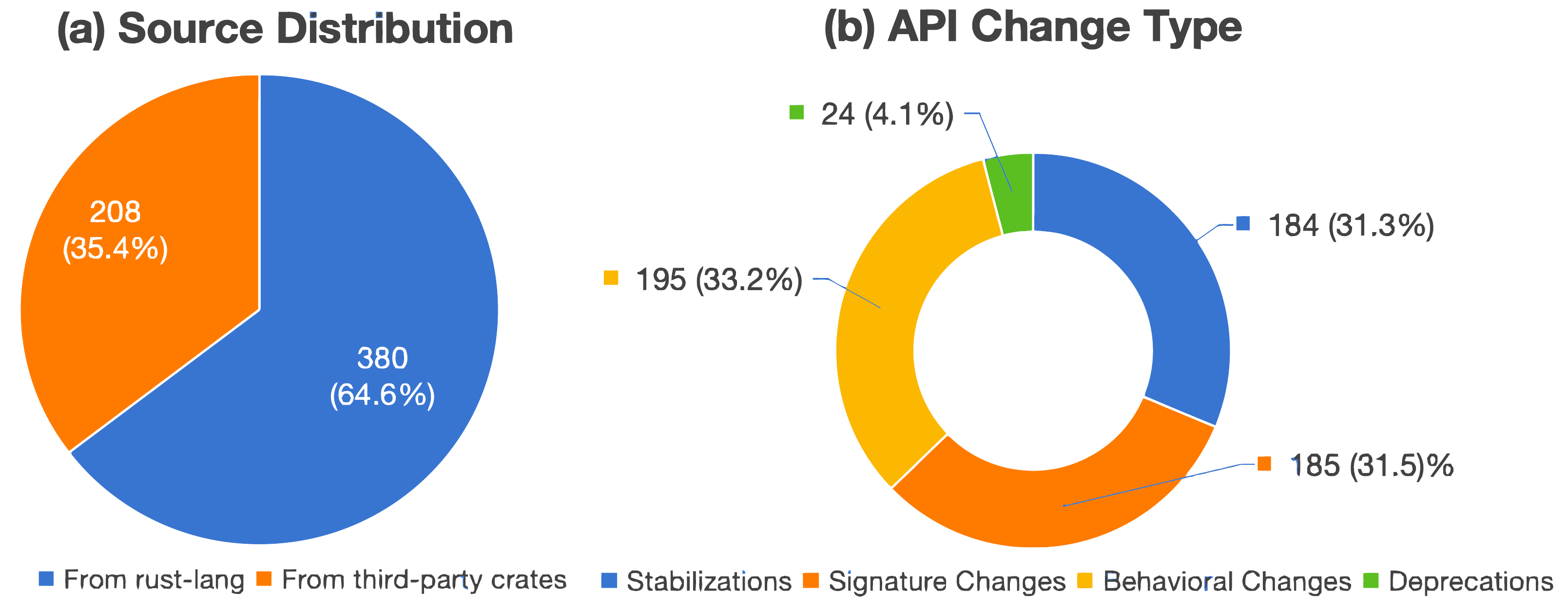}
    \caption{Distribution of 588 Rust APIs: (a) Source Distribution, (b) API Change Type.}
    \label{fig:statistics}
    \vspace{-10pt} 
\end{figure}

During the dataset creation process, our generated test programs achieve a pass rate of 85.7\%.
We retain only the 588 passing examples in our final dataset, which collectively contain 2,574 test cases. Our dataset comprises API changes from Rust official repository and 15 popular third-party crates. For each API, the dataset includes the corresponding programming task, function solution with an average of 17.8 lines of code, and test program designed to verify correct API usage. The wide distribution across different change types allows for an extensive evaluation of model performance in handling various aspects of API evolution. Detailed dataset statistics can be found in Figure~\ref{fig:statistics}.


\subsection{Model Selection}
We select a set of LLMs for our experiments, considering different architectures, parameter scales, and knowledge cutoff dates:

\textbf{For RQ1, RQ2 and RQ5:} We evaluate ten models as representatives of SOTA LLMs with different architectures and training approaches. We include both closed-source models and open-source models to provide a balanced comparison. These models were selected based on their prominence in the field, recency of release, and varying knowledge cutoff dates relative to Rust version releases.

\textbf{For RQ3}: We organize models by their knowledge cutoff dates, focusing on those with cutoff dates before and after specific Rust version releases. This allows us to analyze how temporal knowledge boundaries affect performance on API evolution tasks.

\textbf{For RQ4}: We select the best-performing models from RQ1 for our RAG experiments, enabling us to evaluate whether retrieval augmentation can effectively address knowledge limitations. We implement several retrieval strategies using our API evolution documentation as the knowledge base.

For all models, we use consistent prompt formats and temperature settings (0.7) to guarantee fair comparisons. Each model generates solutions to the same set of tasks from \bench.

\subsection{Metrics}

We adopt the following metrics to evaluate model performance on our benchmark:

\begin{itemize}[leftmargin=1.0em, labelsep=0.5em, itemsep=0.5em, topsep=0.5em] 
    \item \textbf{Pass@k:} The percentage of tasks for which at least one correct solution appears in the top k generations. We primarily focus on Pass@1, which represents the model's capability to generate a correct solution on the first attempt. A task is considered passed only when the generated solution successfully executes all associated test cases for the task.
    \item  \textbf{API Usage Accuracy:} The percentage of generated solutions that correctly implement the required API according to static analysis. We employ different verification methods for each API change type. For Stabilizations, we check if the code correctly uses the newly stabilized API. For Signature Changes, we verify that the code adapts to modified parameters, return types, or generic constraints introduced in the updated API. For Behavioral Changes, we analyze whether the code accounts for the modified behavior. For Deprecations, we confirm that the code uses the recommended alternative APIs rather than deprecated ones. 
    \item \textbf{Test Coverage:} The percentage of test cases that the generated code solution passes. This measures how completely the solution satisfies the requirements beyond just syntactic correctness, capturing whether the code properly handles various inputs and edge cases as defined in our test suite. Unlike Pass@1 which is binary (pass/fail) at the task level, Test Coverage provides a more granular measure by calculating the proportion of individual test cases passed across all tasks.
\end{itemize}

\section{RESULTS}
In this section, we present the results of our experiments, addressing each research question in turn.

\subsection{RQ1: Model Performance Comparison}

We evaluate 10 SOTA LLMs on our \bench benchmark to assess their ability to handle evolving Rust APIs, where each model is provided with the corresponding API information during testing. Table~\ref{tab:model-performance} presents the Pass@1 rates, API Usage Accuracy (AUA), and Coverage percentages for each model across all programming tasks in our dataset.

\begin{table}[h]
    \centering
    \caption{Overall performance of different models on \bench}
    \label{tab:model-performance}
    \setlength{\tabcolsep}{3.0pt}
    \begin{tabular}{lccc}
    \toprule
    \textbf{Model} & \textbf{Pass@1 (\%)} &  \textbf{AUA (\%)} & \textbf{Coverage (\%)} \\
    \midrule
    Claude-3.7-Sonnet &\textbf{65.3}  &\textbf{78.2} &83.6  \\
    o1-mini               &57.5      &70.4    &\textbf{85.2} \\
    GPT-4o                &55.4      &68.4    &77.2   \\
    Gemini-1.5-Pro        &55.3      &62.6    &60.9   \\
    DeepSeek-v3           &54.8      &69.7    &71.0   \\
    Gemini-2.0-Flash      &52.6      &73.5    &72.5   \\
    Llama-3.1-70B         &51.0      &65.3    &69.0   \\
    Qwen-2.5-72B &50.9      &66.7    &64.7   \\
    Claude-3.5-Sonnet     &48.1      &68.7    &80.3   \\
    Grok-3                &40.5      &67.2    &70.4   \\
    \bottomrule
    \end{tabular}
    \vspace{-5pt}
\end{table}

The results show Claude-3.7-Sonnet achieves the highest Pass@1 rate of 65.3\% and AUA of 78.2\%, generating correct code implementations for nearly two-thirds of all Rust API tasks on the first attempt.
Interestingly, Grok-3 exhibits a relatively lower Pass@1 rate of 40.5\% but maintains high AUA (67.2\%) and Coverage (70.4\%), suggesting it understands which APIs to use but struggles with complete implementation details. 



\begin{tcolorbox}[
  enhanced,
  colback=gray!10,      
  colframe=gray!10,      
  boxrule=0.5pt,         
  left=3pt,right=3pt,top=3pt,bottom=3pt,  
  leftrule=4pt,           
  toprule=0.5pt,         
  bottomrule=0.5pt,       
  rightrule=0.5pt,       
  arc=0mm,                
  leftupper=0mm,          
  width=\linewidth,       
  enhanced jigsaw,        
  overlay={
    \fill[black] ([xshift=-4pt]frame.north west) rectangle (frame.south west);
  }
]
\textbf{Answer to RQ1:} 
We observe performance gaps across models on Rust API evolution tasks, with top models achieving 65.3\% Pass@1 rates while others struggle below 50\%. This highlights the challenge LLMs face in adapting to evolving APIs, with newer models generally demonstrating better capabilities for handling Rust's rapidly changing ecosystem. 
\end{tcolorbox}

\subsection{RQ2: Performance Across API Change Types}

To understand how models handle different types of API evolution, we analyze performance across the four categories of API changes in our dataset: Stabilizations (Stab.), Signature Changes (Sig.), Behavioral Changes (Behav.), and Deprecations (Depr.). Table~\ref{tab:api_change_performance} shows the Pass@1 rates for each model across these categories.

\begin{table}[h]
\caption{Performance (Pass@1 \%) of models across different API change types.}
\label{tab:api_change_performance}
\setlength{\tabcolsep}{2.1pt}
\centering
\begin{tabular}{lcccc}
\toprule
\textbf{Model} & \textbf{Stab. (\%)} & \textbf{Sig. (\%)} & \textbf{Behav. (\%)} & \textbf{Depr. (\%)} \\
\midrule
Claude-3.7-Sonnet & \textbf{78.3} & \textbf{72.4} & 45.1 & \textbf{75.0} \\
o1-mini & 71.2  & 67.0 & 37.9 & 37.5  \\
GPT-4o & 65.8 & 65.9 & 38.5  & 33.3  \\
Gemini-1.5-Pro & 69.6  & 55.7  & 40.5 & 62.5  \\
DeepSeek-v3 & 61.4 & 60.0  & \textbf{47.7}  & 20.8  \\
Gemini-2.0-Flash & 58.7  & 69.7  & 32.8  & 33.3 \\
Llama-3.1-70B & 67.4 & 49.7  & 36.4  & 54.2 \\
Qwen-2.5-72B & 63.0 & 56.8 & 36.9  & 25.0  \\
Claude-3.5-Sonnet & 59.8  & 51.9  & 35.9  & 29.2 \\
Grok-3 & 62.5 & 32.4  & 28.2  & 33.3 \\
\midrule
\textbf{Average} & 65.8 & 58.2 & 38.0 & 40.4 \\
\bottomrule
\end{tabular}
\vspace{-8pt}
\end{table}

The performance disparities across API change types reveal limitations in how LLMs process and reason about API evolution. While the average performance follows a clear pattern (Stab. > Sig. > Depr. > Behav.), individual models show interesting variations in their strengths and weaknesses.
Most models perform best on stabilized APIs, with an average Pass@1 rate of 65.8\%. 
Behavioral changes consistently present the greatest challenge for all models, with only a 38.0\% average Pass@1 rate. This represents a substantial performance gap compared to stabilized APIs for most models. Notably, DeepSeek-v3 achieves the highest performance on behavioral changes (47.7\%), significantly outperforming even top-tier models in this challenging category, which suggests its architecture may be better suited for detecting implicit functional modifications.

Our findings reveal that while general trends exist across models, individual models demonstrate unique strengths with specific change types. 


\begin{tcolorbox}[
  enhanced,
  colback=gray!10,      
  colframe=gray!10,      
  boxrule=0.5pt,         
  left=3pt,right=3pt,top=3pt,bottom=3pt,  
  leftrule=4pt,           
  toprule=0.5pt,         
  bottomrule=0.5pt,       
  rightrule=0.5pt,       
  arc=0mm,                
  leftupper=0mm,          
  width=\linewidth,       
  enhanced jigsaw,        
  overlay={
    \fill[black] ([xshift=-4pt]frame.north west) rectangle (frame.south west);
  }
]
\textbf{Answer to RQ2:} 
Model performance varies across API change types, with most models performing best on stabilized APIs (67.4\% average) and all models struggling with behavioral changes (42.1\% average). While some models show unique strengths, the consistent difficulty with behavioral changes highlights a fundamental limitation in detecting semantic modifications without signature alterations.
\end{tcolorbox}

\subsection{RQ3: Knowledge Evolution and Model Performance}

To understand the impact of knowledge cutoff dates, we conduct a controlled experiment that examines how effectively models handle API changes released before versus after their respective training cutoff points. We carefully divided our dataset into two subsets:
(1) Before-cutoff tasks (204 examples): Built using Rust APIs from versions 1.71.0-1.73.0 (July 2023 - October 2023), representing changes that occurred before the earliest model's knowledge cutoff date;
(2) After-cutoff tasks (220 examples): Built using Rust APIs from versions 1.81.0-1.84.0 (September 2024 - January 2025), representing changes that occurred after all models' knowledge cutoff dates. 
For each subset, we test models under two conditions: with and without explicit API information provided. Table~\ref{tab:cutoff_performance} presents the Pass@1 rates for each models across these experimental conditions.

Models perform consistently better on tasks involving API changes before their cutoff dates, achieving average Pass@1 rates of 70.9\% with explicit API information and 56.1\% without. In contrast, performance drops for after-cutoff tasks, averaging 41.4\% with information and 32.5\% without. 

Providing API information increases the performance gap between before-cutoff and after-cutoff APIs, from an average of 23.6\% (56.1\% - 32.5\%) to 29.5\% (70.9\% - 41.4\%). This suggests that models leverage API information more effectively for APIs they were trained on, enabling them to generate correct code more reliably. For after-cutoff APIs, however, models may struggle to apply the information due to a lack of prior knowledge, possibly leading to confusion about API versions.

Interestingly, models still achieve approximately 30\% Pass@1 on after-cutoff tasks without prior knowledge or API information. This performance likely stems from their ability to infer functionality from API names or apply knowledge from similar pre-cutoff APIs.

\begin{table}
    \centering
    \caption{Performance (Pass@1 \%) on API changes before and after model knowledge cutoff dates}
    \label{tab:cutoff_performance}
    \setlength{\tabcolsep}{2.1pt}
    \begin{tabular}{ll|cc|cc}
    \toprule
    \multirow{2}{*}{\textbf{Model}} & \multirow{2}{*}{\textbf{Cutoff}} & \multicolumn{2}{c|}{\textbf{Before Cutoff API}} & \multicolumn{2}{c}{\textbf{After Cutoff API}} \\
    \cline{3-6}
    & & \textbf{w/ Info} & \textbf{w/o Info} & \textbf{w/ Info} & \textbf{w/o Info} \\
    \midrule
    DeepSeek-v3         &2024.07  &71.6  &57.4  &\textbf{46.4}  &34.1   \\
    Gemini-2.0-Flash    &2024.06  &73.5  &54.4  &40.0  &\textbf{35.9}   \\
    Gemini-1.5-Pro      &2024.03   &69.1  &52.9  &42.7  &32.7   \\
    Llama-3.1-70B       &2023.12  &67.2  &53.9  &36.4  &29.1   \\
    o1-mini             &2023.10  &\textbf{73.0}  &\textbf{61.8}  &41.4  &30.5   \\
    \midrule
    \textbf{Average} & - &70.9  &56.1  &41.4  &32.5  \\
    \bottomrule
    \end{tabular} 
    \vspace{-15pt} 
\end{table}


\begin{tcolorbox}[
  enhanced,
  colback=gray!10,      
  colframe=gray!10,      
  boxrule=0.5pt,         
  left=3pt,right=3pt,top=3pt,bottom=3pt,  
  leftrule=4pt,           
  toprule=0.5pt,         
  bottomrule=0.5pt,       
  rightrule=0.5pt,       
  arc=0mm,                
  leftupper=0mm,          
  width=\linewidth,       
  enhanced jigsaw,        
  overlay={
    \fill[black] ([xshift=-4pt]frame.north west) rectangle (frame.south west);
  }
]
\textbf{Answer to RQ3:} 
Knowledge cutoff dates impact model performance on tasks involving Rust API changes. Models perform better when handling APIs within their prior knowledge (56.1\%) compared to those outside their training scope (32.5\%). 
\end{tcolorbox}

\subsection{RQ4: RAG Method Effectiveness}
To evaluate the effectiveness of RAG in improving model performance on API evolution tasks, we compare three approaches: baseline performance without API information, RAG-enhanced generation using API evolution documentation, and direct provision of ground truth API information. For our RAG implementation, we use LangChain as the framework \cite{topsakal2023creating}. Within this framework, we employ OpenAI's text-embedding-3-large as the embedding model to encode API documentation for similarity-based retrieval, while leveraging GPT-4o-mini as the chat model to summarize the relevant API information retrieved.

Table~\ref{tab:rag-impact} shows the performance of five selected models under these three conditions. Without relevant API information, models struggle with recently released APIs, achieving only 34.9\% accuracy on average. When enhanced with RAG, performance improve substantially to 48.4\%. Direct provision of API information yields the best results at 57.7\% on average. 
Claude-3.7-Sonnet demonstrate the strongest performance across all conditions, suggesting it superior capability for reasoning about and adapting to API changes. This may be attributed to its more recent knowledge cutoff date or more comprehensive training on API documentation.

The improvement from RAG (+13.5\% on average) shows that dynamic knowledge retrieval can substantially mitigate the challenges of outdated model knowledge. However, the remaining gap between RAG and direct API information indicates the inherent challenge of API evolution that our benchmark successfully captures. This confirms the necessity of our \bench as a specialized benchmark for evaluating model adaptation to API changes.

\begin{table}[h]
\centering
    \caption{Impact of RAG on model performance for APIs released after knowledge cutoff}
    \label{tab:rag-impact}
        \begin{tabular}{lcccc}
            \toprule
            \textbf{Model} & \textbf{w/o API Info} & \textbf{w/ RAG} & \textbf{w/ API Info} \\
            \hline
            Claude-3.7-Sonnet & \textbf{41.2} & \textbf{54.8} & \textbf{65.3} \\
            o1-mini           & 35.6 & 49.3 & 57.5 \\
            DeepSeek-v3       & 34.0 & 46.5 & 54.8 \\
            Gemini-1.5-Pro    & 32.7 & 42.9 & 55.3 \\
            GPT-4o            & 30.8 & 48.4 & 55.4 \\
            \midrule
            \textbf{Average}  & 34.9 & 48.4 & 57.7 \\
            \bottomrule
        \end{tabular}
        \vspace{-5pt}
\end{table}


\begin{tcolorbox}[
  enhanced,
  colback=gray!10,      
  colframe=gray!10,      
  boxrule=0.5pt,         
  left=3pt,right=3pt,top=3pt,bottom=3pt,  
  leftrule=4pt,           
  toprule=0.5pt,         
  bottomrule=0.5pt,       
  rightrule=0.5pt,       
  arc=0mm,                
  leftupper=0mm,          
  width=\linewidth,       
  enhanced jigsaw,        
  overlay={
    \fill[black] ([xshift=-4pt]frame.north west) rectangle (frame.south west);
  }
]
\textbf{Answer to RQ4:}
RAG substantially improves model performance on API evolution tasks, bridging approximately 60\% of the gap between no API information and complete API information. This confirms RAG as an effective approach for keeping code generation models current with evolving APIs without requiring complete retraining. 
\end{tcolorbox}

\subsection{RQ5: Model Performance on \bench vs. Stable APIs}
\begin{figure}
    \centering
    \includegraphics[width=1.0\linewidth]{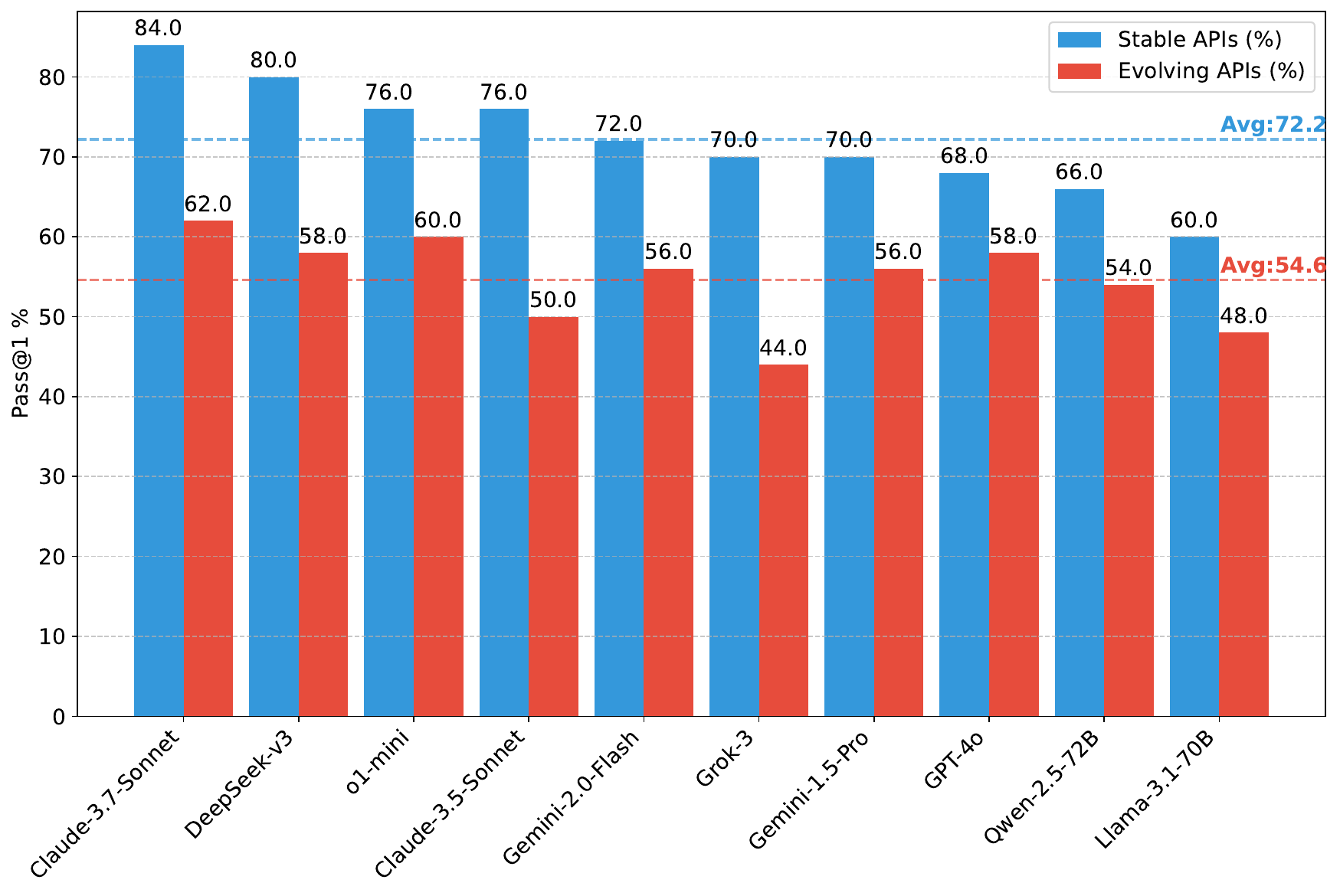}
    \caption{Performance comparison (Pass@1 \%) on evolving APIs vs. stable APIs}
    \label{fig:rq5}
\end{figure}

To quantify the specific challenges introduced by API evolution, we compare model performance on our \bench benchmark against a control dataset of 50 stable APIs from the official Rust repository. 
Figure~\ref{fig:rq5} presents the Pass@1 results across 10 models. The results reveal a substantial performance decline between stable and evolving APIs across all models. On average, models achieve 72.2\% Pass@1 on stable APIs compared to only 54.6\% on evolving APIs, representing a 17.6 percentage point drop. This performance degradation is consistently observed across all tested models regardless of their their architecture or size. 


Claude-3.7-Sonnet demonstrates the strongest performance on both stable (84.0\%) and evolving APIs (62.0\%), but still exhibits a 22.0 percentage point gap between the two task types. Interestingly, while DeepSeek-v3 achieves the second-highest score on stable APIs (80.0\%), o1-mini performs better on evolving APIs (60.0\% vs. 58.0\%), suggesting that certain models may be more resilient to API changes. The performance drop is most pronounced for Grok-3, which experiences a 26.0 percentage point decrease (70.0\% to 44.0\%), indicating particular difficulty adapting to API evolution.

This performance gap across all models confirms that API evolution introduces specific challenges beyond the inherent complexity of programming tasks. Even when controlling for source repository and task construction methodology, the evolving nature of Rust APIs impacts model performance, requiring specialized capabilities to detect and adapt to behavioral changes.


\begin{tcolorbox}[
  enhanced,
  colback=gray!10,      
  colframe=gray!10,      
  boxrule=0.5pt,         
  left=3pt,right=3pt,top=3pt,bottom=3pt,  
  leftrule=4pt,           
  toprule=0.5pt,         
  bottomrule=0.5pt,       
  rightrule=0.5pt,       
  arc=0mm,                
  leftupper=0mm,          
  width=\linewidth,       
  enhanced jigsaw,        
  overlay={
    \fill[black] ([xshift=-4pt]frame.north west) rectangle (frame.south west);
  }
]
\textbf{Answer to RQ5:} 
Our experiment shows a consistent performance drop with evolving APIs compared to stable ones across all models, highlighting the challenges of Rust API evolution beyond general programming complexity and the need for evolution-aware benchmarks like \bench.
\end{tcolorbox}

\section{THREATS TO VALIDITY}


\textbf{Task Generation} While our LLM-based task generation pipeline produces realistic programming tasks, it may introduce biases from the underlying model. DeepSeek-v3's own limitations could affect the variety and complexity of generated tasks. Additionally, tasks generated by LLMs may be easier for other LLMs to solve than real-world programming challenges, potentially leading to inflated performance metrics. We mitigate this by validating tasks with executable test programs and using realistic usages as seed content.

\noindent \textbf{Third-Party Selection} The selection of third-party crates, while based on quantitative criteria, may not fully represent the broader Rust ecosystem. Popular crates may have better documentation and more stable API evolution patterns than the average crate, potentially making our tasks less representative of challenges faced with less-maintained libraries. Our approach also focuses on crates with adequate documentation, which excludes libraries with limited or outdated documentation despite possible widespread usage.

\noindent \textbf{Model Evaluation} Our evaluation primarily focuses on Pass@1 metrics using our test programs. While these programs verify functional correctness, they may not capture all aspects of code quality such as efficiency, readability, or adherence to idiomatic Rust practices. Additionally, we test a limited set of models, and our findings may not generalize to all LLMs or future model architectures. The RAG implementation we used represents just one possible approach, and alternative retrieval strategies might yield different results.


\section{RELATED WORK}

\subsection{API Evolution}

API evolution has been widely studied across programming languages ~\cite{10.1145/3639478.3640041, MAHMUD2023111664, 10.1007/978-3-642-03013-0_15}, with a focus on Java due to its widespread use in enterprise and mobile ecosystems\cite{lamothe2021systematic}. For instance, Dagenais and Robillard introduce SemDiff, a tool to automate API migration by analyzing call dependencies in Java frameworks, highlighting challenges in maintaining backward compatibility during API changes ~\cite{dagenais2009semdiff}. McDonnell et al. examine the Android API’s rapid evolution, noting that 28\% of API references in client applications became outdated, with migration delays averaging 14 months ~\cite{mcdonnell2013empirical}. Java’s dominance in API evolution research is further evident in studies on API deprecation patterns, where Brito et al. find that 64\% of deprecated Java APIs lacked adequate replacement documentation ~\cite{brito2018use}.



\textbf{Rust-specific studies} are notably absent in the surveyed literature. While Rust’s ownership system and memory safety features introduce distinct API design constraints, no prior work explicitly addresses its evolution challenges.

\subsection{Code Generation Evaluation}




Prior work on code generation evaluation falls into version-agnostic and API-version-aware approaches.

Traditional methods~\cite{austin2021program, lu2021codexglue, husain2019codesearchnet, quan2025codeelo, li2024evocodebench, NEURIPS2023_43e9d647} assess generated code using syntactic similarity metrics like BLEU \cite{papineni2002bleu} or unit test pass rates \cite{chen2021evaluating}. However, they lack API version specification, leading to misleading evaluations. Models may use deprecated APIs that pass tests, while correct use of newer features may be penalized, discouraging adoption of updated APIs and modern language features.

Recent work addresses this limitation through API-version-aware evaluation frameworks. Versicode ~\cite{wu2024versicode} introduces manually verified test cases to assess API function implementation in LLM-generated code. Wang~et~al. ~\cite{wang2025llmsmeetlibraryevolution} propose a classification scheme that categorizes API usage as deprecated, replacement, or non-standard. CodeUpdateArena~\cite{liu2024codeupdatearena} evaluates model adaptation through synthetic API updates, testing whether edited models can apply new APIs to specific program synthesis tasks.


\subsection{Code Generation Datasets Construction}



Code generation datasets pair natural language prompts with code snippets for training and evaluating LLMs. Their construction methods fall into two categories: real-world data collection, where humans gather and annotate data, and LLM-assisted construction, where models generate or annotate content.

\textbf{Real-World Data Collection.} Real-world data collection involves gathering code from existing sources. This category is further divided into three sub-methods. (1) Collecting from competitive programming platforms. For example, CodeContests ~\cite{li2022competition} and APPS ~\cite{hendrycks2021measuring} comprise coding problems collected from open-access coding websites. Their construction process involves gathering problems, associated test cases and human solutions. (2) Collecting from real-world projects. Collecting code from realistic projects, often from platforms like GitHub, can reflect practical and diverse coding scenarios. For instance, the Stack ~\cite{kocetkov2022stack} employs rule-based filtering of open-source repositories, ultimately containing over 6TB of permissively licensed source code from GitHub. (3) Manually creating problems. This method is time-consuming and may not scale well for large datasets. For instance, HumanEval is a hand-crafted dataset consisting of 164 programming problems, each including a function signature, docstring, code body, and unit tests~\cite{chen2021evaluating}.

\textbf{LLM-Assisted Construction.} LLMs are increasingly used to generate or augment datasets for code generation tasks, providing a scalable and efficient way to create high-quality data. Recent advances in LLMs enable four primary approaches for fully automated or semi-automated dataset creation. (1) Code Synthesis and Generation. LLMs can directly generate synthetic code samples paired with natural language descriptions, such as Codex ~\cite{chen2021evaluating} and Code Llama ~\cite{roziere2023code}. (2) Instruction Generation. LLMs can generate programming problems and instructional text through prompt engineering. For example, VersiCode generates function descriptions for version-specific code completion ~\cite{wu2024versicode}. (3) Automated Annotation and Evaluation. LLMs can create test suites for code samples ~\cite{liu2024codeupdatearena, gong2025cosqapioneeringmultichoicecode}. LLMs can also rank solutions to filter optimal implementations ~\cite{shypula2023perfect}.

\section{CONCLUSION}

We present \bench, a two-phase framework for evaluating LLMs' ability to handle Rust API evolution. Our approach collects API changes from multiple sources and transforms them into natural programming tasks. The resulting dataset covers 588 API changes across four categories spanning Rust versions 1.71.0-1.84.0. Experiments show models perform best on stabilizations (65.8\% average Pass@1) and worst on behavioral changes (38.0\%), with performance declining as temporal distance from knowledge cutoff dates increased. RAG provided substantial improvements (+13.5\%) for APIs released after model cutoffs. By releasing our framework and dataset, we aim to facilitate research into API-aware code generation for evolving programming interfaces.


\bibliographystyle{ACM-Reference-Format}
\bibliography{ref}

\end{document}